\documentclass[referee]{aa} 
\input{psfig.sty}

\begin{document}

\title{Doppler effect of gamma-ray bursts in the fireball framework}

\author{Yi-Ping Qin\inst{1,2}\fnmsep\thanks{e-mail:ypqin@public.km.yn.cn}
           \and
           Fu-Wen Zhang\inst{3}
          }

\offprints{Yi-Ping Qin}

\institute{Yunnan Observatory, National Astronomical
Observatories, CAS, Kunming, Yunnan 650011, P. R. China \and
Chinese Academy of Science-Peking University joint Beijing
Astrophysical Center,  P. R. China \and Physics Department,
Guangxi University, Nanning, Guangxi, 530004, P. R. China}
\date{Received                  ; accepted                }

\abstract{ The influence of the Doppler effect in the fireball
framework on the spectrum of gamma-ray bursts is investigated. The
study shows that the shape of the expected spectrum of an
expanding fireball remains almost the same as that of the
corresponding rest frame spectrum for constant radiations of the
bremsstrahlung, Comptonized, and synchrotron mechanisms as well as
for that of the GRB model. The peak flux spectrum and the peak
frequency are obviously correlated. When the value of the Lorentz
factor becomes 10 times larger, the flux of fireballs would be
several orders of magnitude larger. The expansion speed of
fireballs is a fundamental factor of the enhancement of the flux
of gamma-ray bursts.  }
\titlerunning{Doppler effect in the fireball framework}

\authorrunning{Y.-P. Qin}

\maketitle

\section{Introduction}

Gamma-ray bursts (GRBs) are transient astrophysical phenomena in which the
emission is confined exclusively to high energies: they are detected at
gamma-ray bands and have short lifetimes (from a few milliseconds to several
hundred seconds). Since the discovery of the objects about thirty years ago (%
\cite{Kl73}), many properties have been revealed. At the same
time, various models accounting for the observation have been
proposed. Due to the observed great output rate of radiation, most
models envision an expanding fireball (see e.g., \cite{Go86};
\cite{Pa86}). The gamma-ray emission would arise after the
fireball becomes optically thin, in shocks produced when the
ejecta collide with an external medium or occurred within a
relativistic internal wind (Rees \& Meszaros 1992, 1994; Meszaros
\& Rees 1993, 1994; \cite{Ka94}; \cite{Pa94}; \cite{Sa96}). As
there might be a strong magnitude field within the fireball and
the expansion would be relativistic, it was
believed that the synchrotron radiation would become a dominate mechanism (%
\cite{Ra81}; \cite{Li83}). Unfortunately, the spectra of the
objects are so different that none of the mechanisms proposed so
far can account for most observed data (\cite{Ba93}; \cite{Sc94};
Preece et al. 1998, 2000).

{\bf As pointed out by Krolik \& Pier (1991), relativistic bulk
motion of the gamma-ray-emitting plasma can account for some
phenomena of bursts. However, in some cases, the whole fireball
surface should be considered.} As the expanding motion of the
outer shell of the fireball would be relativistic, the Doppler
effect must be at work and in considering the effect the fireball
surface itself must play a role (\cite{Me98}). In the following we
will present a detailed study of the effect and then analyse the
influenced spectrum of fireballs radiating under various
mechanisms.

\section{Doppler effect in the fireball framework}

As the first step of investigation, we concern in the following only the
core content of the Doppler effect in the fireball framework, with the
cosmological as well as other effects being temporarily ignored, and pay our
attention only to the fireball expanding at a constant velocity $v=\beta c$
(where $c$ is the speed of light) relative to its center.

Let the distance between the observer and the center of the fireball be $D$,
with $D\gg $ the radius of the fireball. Suppose photons from the rest frame
differential surface, $ds_{0,\theta ,\varphi }$, of the fireball arriving
the observer at time $t$ are emitted at proper time $t_{0,\theta }$, {\bf %
where }$\theta ${\bf \ denotes the angle to the line of sight and }$%
\varphi ${\bf \ denotes the other angular coordinate} {\bf of the
fireball surface}. Let $ds_{\theta ,\varphi }$ be the
corresponding
differential surface resting on the observer framework, coinciding with $%
ds_{0,\theta ,\varphi }$ at $t_{0,\theta }$, and $t_\theta $ be the
corresponding coordinate time of the moment $t_{0,\theta }$. We assign $%
t_\theta =t_c$ when $t_{0,\theta }=t_{0,c}$, where $t_c$ and $t_{0,c}$ are
constants. Considering the travelling of light from the fireball to the
observer, one can obtain the following relations (see Appendix A):
\begin{equation}
t_\theta =\frac{t-D/c+(R_c/c-\beta t_c)\cos \theta }{1-\beta \cos \theta },
\end{equation}
\begin{equation}
t_{0,\theta }=\frac{t-t_c-D/c+(R_c/c)\cos \theta }{\Gamma (1-\beta \cos
\theta )}+t_{0,c},
\end{equation}
and
\begin{equation}
R[t_\theta (t)]=R_0[t_{0,\theta }(t)]=\frac{\widetilde{R}(t)}{1-\beta \cos
\theta },
\end{equation}
with
\begin{equation}
\widetilde{R}(t)=\beta [c(t-t_c)-D]+R_c,
\end{equation}
where{\bf \ }$\Gamma ${\bf \ is the Lorentz factor}, $R(t_\theta
)$
[or $R_0(t_{0,\theta })$] is the radius of the fireball at $t_\theta $ [or $%
t_{0,\theta }$], and $R_c$ is the radius at time $t_\theta =t_c$ (or $%
t_{0,\theta }=t_{0,c}$).

For a radiation independent of directions, when considering the observed
amount of energy emitted from the whole object surface, one can obtain the
following flux expected from the expanding fireball:
\begin{equation}
f_\nu (t)=\frac{2\pi }{D^2}\int_{\theta _{\min }}^{\theta _{\max
}}R^2(t_\theta )I_\nu (t_\theta ,\nu )\cos \theta \sin \theta d\theta ,
\end{equation}
where $I_\nu (t_\theta ,\nu )$ is the observer frame intensity. Replacing $%
I_\nu (t_\theta ,\nu )$ with the rest frame intensity $I_{0,\nu
}(t_{0,\theta },\nu _{0,\theta })$, where {\bf the rest frame frequency }$%
\nu _{0,\theta }${\bf \ is related with the observation frequency
}$\nu $ by the Doppler effect, one can obtain the following form
for the expected flux:
\begin{equation}
f_\nu (t)=\frac{2\pi }{D^2\Gamma ^3}\int_{\theta _{\min }}^{\theta _{\max }}%
\frac{R_0^2(t_{0,\theta })I_{0,\nu }(t_{0,\theta },\nu _{0,\theta })\cos
\theta \sin \theta }{(1-\beta \cos \theta )^3}d\theta .
\end{equation}

The integral limits of $\theta $ in (6) should be determined by the emitted
ranges of t$_{0,\theta }$ and $\nu _{0,\theta }$ together with the fireball
surface itself. Let
\begin{equation}
t_{0,\min }\leq t_{0,\theta }\leq t_{0,\max }
\end{equation}
and
\begin{equation}
\nu _{0,\min }\leq \nu _{0,\theta }\leq \nu _{0,\max }.
\end{equation}
We find that, when the following condition
\begin{equation}
\max \{\theta _{t,\min },\theta _{\nu ,\min }\}<\min \{\theta _{t,\max
},\theta _{\nu ,\max }\}
\end{equation}
is satisfied, $\theta _{\min }$ and $\theta _{\max }$ would be determined by
\begin{equation}
\theta _{\min }=\max \{\theta _{t,\min },\theta _{\nu ,\min }\}
\end{equation}
and
\begin{equation}
\theta _{\max }=\min \{\theta _{t,\max },\theta _{\nu ,\max }\},
\end{equation}
respectively, where
\begin{equation}
\begin{tabular}{c}
$\theta _{t,\min }=\cos ^{-1}\left( \min \{1,\frac{\Gamma
c(t_{0,\max }-t_{0,c})+D-c(t-t_c)}{\Gamma \beta c(t_{0,\max
}-t_{0,c})+R_c}\}\right) $
\\
$\left( 0\leq \frac{\Gamma c(t_{0,\max
}-t_{0,c})+D-c(t-t_c)}{\Gamma \beta
c(t_{0,\max }-t_{0,c})+R_c}\right) ,$%
\end{tabular}
\end{equation}
\begin{equation}
\begin{tabular}{c}
$\theta _{t,\max }=\cos ^{-1}\left( \max \{0,\frac{\Gamma
c(t_{0,\min }-t_{0,c})+D-c(t-t_c)}{\Gamma \beta c(t_{0,\min
}-t_{0,c})+R_c}\}\right) $
\\
$\left( \frac{\Gamma c(t_{0,\min }-t_{0,c})+D-c(t-t_c)}{\Gamma
\beta
c(t_{0,\min }-t_{0,c})+R_c}\leq 1\right) ,$%
\end{tabular}
\end{equation}
\begin{equation}
\begin{tabular}{c}
$\theta _{\nu ,\min }=\cos ^{-1}\left( \min \{1,\frac 1\beta (1-\frac{\nu
_{0,\min }}{\Gamma \nu })\}\right) $ \\
$\left( 0\leq \frac 1\beta (1-\frac{\nu _{0,\min }}{\Gamma \nu });0<\beta
\right) ,$%
\end{tabular}
\end{equation}
and
\begin{equation}
\begin{tabular}{c}
$\theta _{\nu ,\max }=\cos ^{-1}\left( \max \{0,\frac 1\beta (1-\frac{\nu
_{0,\max }}{\Gamma \nu })\}\right) $ \\
$\left( \frac 1\beta (1-\frac{\nu _{0,\max }}{\Gamma \nu })\leq 1;0<\beta
\right) .$%
\end{tabular}
\end{equation}
Note that, all the conditions in (12)---(15) and the condition of (9) must
be satisfied, otherwise no emission from the fireball will be detected at
frequency $\nu $ and at time $t$.

For a constant radiation of a continuum, which covers the entire frequency
band, one would get $\theta _{t,\min }=\theta _{\nu ,\min }=0$ and $\theta
_{t,\max }=\theta _{\nu ,\max }=\pi /2$, thus $\theta _{\min }=0$ and $%
\theta _{\max }=\pi /2$. In practice, radiation lasting a sufficient
interval of time would lead to $\theta _{t,\min }=0$ and $\theta _{t,\max
}=\pi /2$, especially when $t_{0,\min }$ and $t_{0,\max }$ are far beyond
the interval [$(ct-ct_c-D)/\Gamma c+t_{0,c},(ct-ct_c-D+R_c)/(\Gamma c-\Gamma
\beta c)+t_{0,c}$] (see Appendix B).

\section{Effect on some continuous spectra}

{\bf Although for some bursts, their time-dependent spectral data
have been published, yet for many GRBs, only average spectral data
are available (in \cite{Pr00}, the number of bursts for which
time-resolved spectra were presented is 156)}. When employed to
fit the average data, constant radiations were always considered
and preferred. In the following we consider constant radiations
and illustrate how the Doppler effect on the fireball model
affecting continuous spectra of some mechanisms.

The bremsstrahlung, Comptonized, and synchrotron radiations were
always taken as plausible mechanisms accounting for gamma-ray
bursts (\cite{Sc94}):
\begin{equation}
I_{0,\nu ,B}(t_{0,\theta },\nu _{0,\theta })=I_{0,B}\exp
(-\frac{\nu _{0,\theta }}{\nu _{0,B}})\quad (-\infty <t_{0,\theta
}<\infty ;0<\nu _{0,\theta }<\infty ),
\end{equation}
\begin{equation}
I_{0,\nu ,C}(t_{0,\theta },\nu _{0,\theta })=I_{0,C}\nu _{0,\theta
}^{1+\alpha _{0,C}}\exp (-\frac{\nu _{0,\theta }}{\nu
_{0,C}})\quad (-\infty <t_{0,\theta }<\infty ;0<\nu _{0,\theta
}<\infty ),
\end{equation}
\begin{equation}
I_{0,\nu ,S}(t_{0,\theta },\nu _{0,\theta })=I_{0,S}\nu _{0,\theta
}\exp [-0.3887(\frac{\nu _{0,\theta }}{\nu _{0,S}})^{1/3}]\quad
(-\infty <t_{0,\theta }<\infty ;0<\nu _{0,\theta }<\infty ),
\end{equation}
where subscripts ``$B$'', ``$C$'' and ``$S$'' represent the bremsstrahlung,
Comptonized and synchrotron radiations, respectively; $\nu _{0,B}=kT_{0,B}/h$
and $\nu _{0,C}=kT_{0,C}/h$, with $T_{0,B}$ and $T_{0,C}$ being the
corresponding temperatures of the bremsstrahlung and Comptonized radiations,
respectively; $\alpha _{0,C}$ is the index of the Comptonized radiation; $%
\nu _{0,S}$ is the synchrotron parameter; $I_{0,B}$, $I_{0,C}$ and $I_{0,S}$
are constants.

Since none of the mechanisms proposed so far can account for most
observed spectral data of bursts, an empirical spectral form
called the GRB model (\cite{Ba93}) was frequently, and rather
successfully, employed to fit most spectra of bursts (see e.g.,
\cite{Sc94}; \cite{Fo95}; Preece et al. 1998, 2000). It is
\begin{equation}
I_{0,\nu ,G}(t_{0,\theta },\nu _{0,\theta })=I_{0,G}g_{0,\nu
,G}(\nu _{0,\theta })\quad (-\infty <t_{0,\theta }<\infty )
\end{equation}
with
\begin{equation}
g_{0,\nu ,G}(\nu _{0,\theta })=\{
\begin{array}{c}
(\frac{\nu _{0,\theta }}{\nu _{0,p}})^{1+\alpha _{0,G}}\exp [-(2+\alpha
_{0,G})\frac{\nu _{0,\theta }}{\nu _{0,p}}]\quad (\frac{%
\nu _{0,\theta }}{\nu _{0,p}}<\frac{\alpha _{0,G}-\beta _{0,G}}{2+\alpha
_{0,G}}) \\
(\frac{\alpha _{0,G}-\beta _{0,G}}{2+\alpha _{0,G}})^{\alpha _{0,G}-\beta
_{0,G}}\exp (\beta _{0,G}-\alpha _{0,G})(\frac{\nu _{0,\theta }}{\nu _{0,p}}%
)^{1+\beta _{0,G}}\quad (\frac{\nu _{0,\theta }}{\nu _{0,p}%
}\geq \frac{\alpha _{0,G}-\beta _{0,G}}{2+\alpha _{0,G}}),
\end{array}
\end{equation}
where subscript ``$G$'' represents the word ``GRB'', ``$p$''
stands for ``peak'', $\alpha _{0,G}${\bf \ and }$\beta _{0,G}${\bf
\ are the lower and higher indexes, respectively,} and $I_{0,G}$
is a constant.

As mentioned in last section, the integral limits of $\theta $ in (6) for
these constant radiations would be $\theta _{\min }=0$ and $\theta _{\max
}=\pi /2$. Applying (3) and (16)---(19) to (6), we get the following
corresponding spectra:
\begin{equation}
\nu f_{\nu ,B}(t)=\frac{2\pi I_{0,B}\widetilde{R}^2(t)\nu _{0,B}}{D^2}\frac
\nu {\Gamma ^3\nu _{0,B}}\int_0^{\pi /2}\exp (-\frac{\nu _{0,\theta }}{\nu
_{0,B}})\frac{\cos \theta \sin \theta }{(1-\beta cos\theta )^5}d\theta ,
\end{equation}
\begin{equation}
\nu f_{\nu ,C}(t)=\frac{2\pi I_{0,C}\widetilde{R}^2(t)\nu _{0,C}^{2+\alpha
_{0,C}}}{D^2}\frac \nu {\Gamma ^3\nu _{0,C}}\int_0^{\pi /2}(\frac{\nu
_{0,\theta }}{\nu _{0,C}})^{1+\alpha _{0,C}}\exp (-\frac{\nu _{0,\theta }}{%
\nu _{0,C}})\frac{\cos \theta \sin \theta }{(1-\beta cos\theta )^5}d\theta ,
\end{equation}
\begin{equation}
\nu f_{\nu ,S}(t)=\frac{2\pi I_{0,S}\widetilde{R}^2(t)\nu _{0,S}^2}{D^2}%
\frac \nu {\Gamma ^3\nu _{0,S}}\int_0^{\pi /2}\frac{\nu _{0,\theta }}{\nu
_{0,S}}\exp [-0.3887(\frac{\nu _{0,\theta }}{\nu _{0,S}})^{1/3}]\frac{\cos
\theta \sin \theta }{(1-\beta cos\theta )^5}d\theta ,
\end{equation}
\begin{equation}
\nu f_{\nu ,G}(t)=\frac{2\pi I_{0,G}\widetilde{R}^2(t)\nu _{0,p}}{D^2}\frac
\nu {\Gamma ^3\nu _{0,p}}\int_0^{\pi /2}g_{0,\nu ,G}(\nu _{0,\theta })\frac{%
\cos \theta \sin \theta }{(1-\beta cos\theta )^5}d\theta ,
\end{equation}
where $\nu $ and $\nu _{0,\theta }$ are related by the Doppler effect, $%
\widetilde{R}(t)$ is shown in (4) and $g_{0,\nu ,G}(\nu _{0,\theta })$ is
shown in (20).

As instances for illustration, typical values, $\alpha _{0,C}=-0.6$
(\cite{Sc94}), for the index of the Comptonized radiation, and $%
\alpha _{0,G}=-1$ and $\beta _{0,G}=-2.25$ (Preece et al. 1998, 2000), for
the lower and higher indexes of the GRB model, are adopted.

For the sake of comparison, we ignore the development of the magnitude of
spectra and consider a particular observation time when photons emitted from
$\theta =\pi /2$ of the fireball with its radius being $R_c$ reach the
observer, which is $t=D/c+t_c$. That leads to $\widetilde{R}(t)=R_c$.

\begin{figure}[tbp]
\vbox to 3.0in{\rule{0pt}{3.0in}} \includegraphics{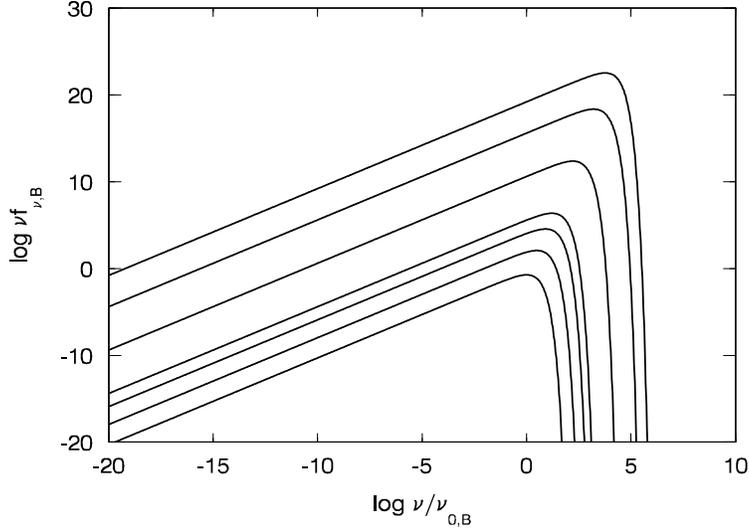} \caption{The
expected spectrum of a fireball with its rest frame radiation
being the bremsstrahlung one, where we take $2\pi $I$_{0,B}\widetilde{R}^2$%
(t)$\nu _{0,B}/$D$^{2}=1$. The solid lines from the bottom to the top
correspond to $\Gamma =$ 1, 2, 5, 10, 100, 1000, and 10000, respectively. }
\label{Fig1}
\end{figure}

\begin{figure}[tbp]
\vbox to 3.0in{\rule{0pt}{3.0in}} \includegraphics{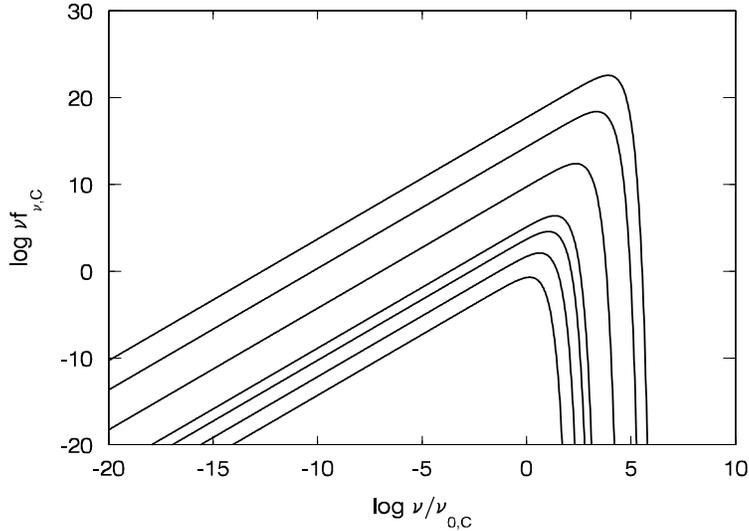} \caption{The
expected spectrum of a fireball with its rest frame radiation
being the Comptonized one, where we take $2\pi $I$_{0,C}\widetilde{R}^2$(t)$%
\nu _{0,C}^{2+\alpha _{0,C}}/$D$^{2}=1$ and $\alpha _{0,C}=-0.6$. The
symbols are the same as those in Fig. 1. }
\label{Fig2}
\end{figure}

\begin{figure}[tbp]
\vbox to 3.0in{\rule{0pt}{3.0in}} \includegraphics{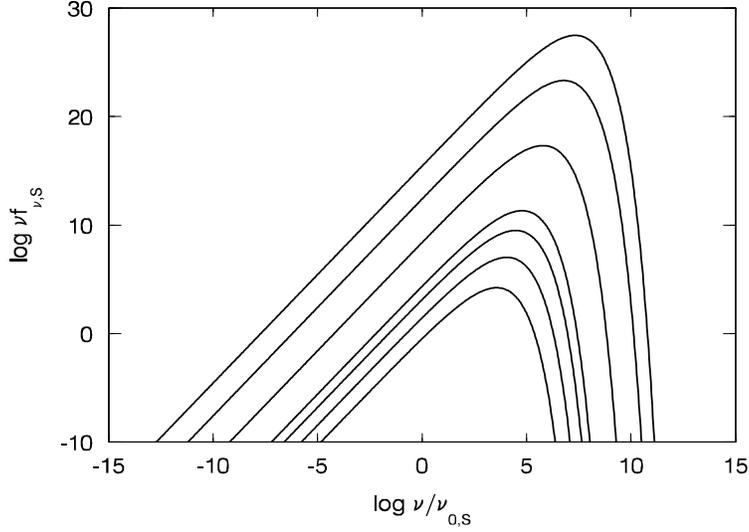} \caption{The
expected spectrum of a fireball with its rest frame radiation
being the synchrotron one, where we take $2\pi $I$_{0,S}\widetilde{R}^2$(t)$%
\nu _{0,S}^2/$D$^{2}=1$. The symbols are the same as those in Fig. 1. }
\label{Fig3}
\end{figure}

\begin{figure}[tbp]
\vbox to 3.0in{\rule{0pt}{3.0in}} \includegraphics{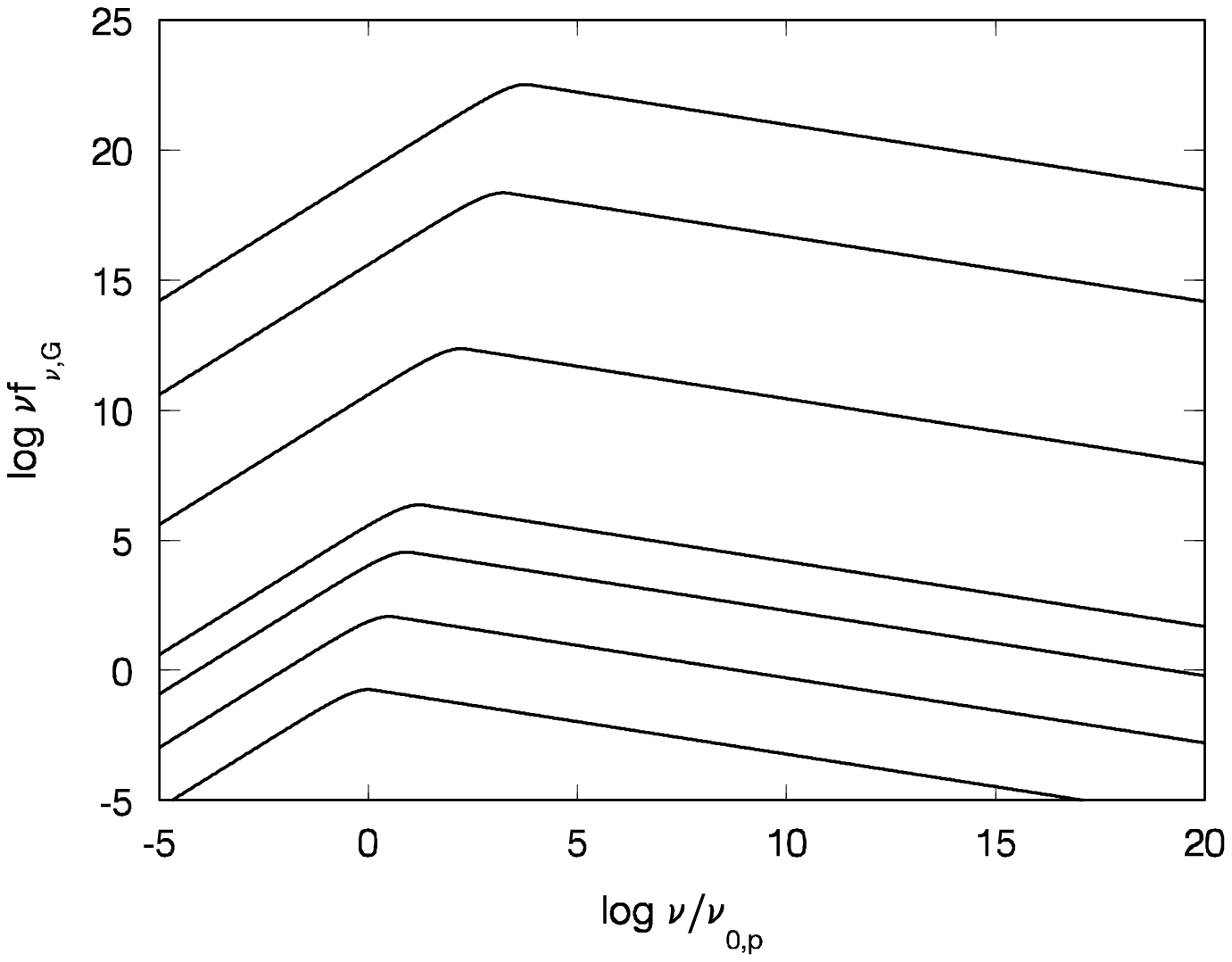} \caption{The
expected spectrum of a fireball with its rest frame radiation
bearing the GRB form, where we take $2\pi
$I$_{0,G}\widetilde{R}^2$(t)$\nu _{0,p}/$D$^{2}=1$, $\alpha
_{0,G}=-1$ and $\beta _{0,G}=-2.25$. The symbols are the same as
those in Fig. 1. } \label{Fig4}
\end{figure}

Shown in Figs. 1--4 are the expected spectra of the bremsstrahlung,
Comptonized, synchrotron and GRB form radiations, respectively, emitted from
a fireball with various values of $\Gamma $ and observed at time $t=D/c+t_c$.

One can find from these figures that the shape of the rest frame spectrum of
the adopted models is not significantly affected by the expansion of
fireballs. However, as the fireball expands, the peak of the spectrum would
shift to a higher frequency band and the flux over the entire frequency
range would be amplified. The enhancement of the flux occurs not only at
higher bands but also at lower bands. Even $\Gamma $ is not very large (say $%
\Gamma =2$), a dim and undetectable X-ray rest frame radiation might become
an observable gamma-ray source.

\section{Relation between the peak frequency and the peak flux spectrum}

As mentioned above, as the fireball expands, the peak of the spectrum would
shift to a higher frequency band and the flux over the entire frequency
range would be amplified. This suggests a correlation between the {\bf %
peak flux spectrum}, $(\nu f_\nu )_p$, and the frequency where this peak is
found, the so called peak frequency $\nu _p$ (or the peak energy $E_p$). To
get a more detail information of this issue, we calculate $\nu _p$ as well
as $(\nu f_\nu )_p$ for some values of $\Gamma $ for the radiations
considered above. The results are listed in Tables 1--4.

\begin{table}[ht]
\caption{List of $\nu _{p}$ and $(\nu $f$_\nu )_{p}$ for the bremsstrahlung
radiation}
\begin{tabular}{lll}
\hline\hline
$\Gamma $ & $\nu _{p}/\nu _{0,B}$ & $(\nu $f$_{\nu ,B})_{p}$ \\ \hline
 $1\times10^{0}$ & $9.99\times10^{-1}$ & $1.84\times10^{-1}$ \\
            $2\times10^{0}$ & $3.06\times10^{0}$ & $1.16\times10^{2}$ \\
            $5\times10^{0}$ & $8.00\times10^{0}$ & $3.48\times10^{4}$ \\
            $1\times10^{1}$ & $1.62\times10^{1}$ & $2.29\times10^{6}$ \\
            $2\times10^{1}$ & $3.23\times10^{1}$ & $1.47\times10^{8}$ \\
            $5\times10^{1}$ & $8.09\times10^{1}$ & $3.60\times10^{10}$ \\
            $1\times10^{2}$ & $1.62\times10^{2}$ & $2.31\times10^{12}$ \\
            $2\times10^{2}$ & $3.25\times10^{2}$ & $1.48\times10^{14}$ \\
            $5\times10^{2}$ & $8.08\times10^{2}$ & $3.60\times10^{16}$ \\
            $1\times10^{3}$ & $1.62\times10^{3}$ & $2.31\times10^{18}$ \\
            $2\times10^{3}$ & $3.24\times10^{3}$ & $1.48\times10^{20}$ \\
            $5\times10^{3}$ & $8.12\times10^{3}$ & $3.60\times10^{22}$ \\
            $1\times10^{4}$ & $1.62\times10^{4}$ & $2.31\times10^{24}$ \\ \hline
\end{tabular}
\end{table}

\begin{table}[ht]
\caption{List of $\nu _{p}$ and $(\nu $f$_\nu )_{p}$ for the Comptonized
radiation}
\begin{tabular}{lll}
\hline\hline
$\Gamma $ & $\nu _{p}/\nu _{0,C}$ & $(\nu $f$_{\nu ,C})_{p}$ \\ \hline
$1\times10^{0}$ & $1.40\times10^{0}$ & $1.97\times10^{-1}$ \\
            $2\times10^{0}$ & $4.30\times10^{0}$ & $1.24\times10^{2}$ \\
            $5\times10^{0}$ & $1.13\times10^{1}$ & $3.71\times10^{4}$ \\
            $1\times10^{1}$ & $2.27\times10^{1}$ & $2.44\times10^{6}$ \\
            $2\times10^{1}$ & $4.56\times10^{1}$ & $1.57\times10^{8}$ \\
            $5\times10^{1}$ & $1.14\times10^{2}$  & $3.84\times10^{10}$ \\
            $1\times10^{2}$ & $2.28\times10^{2}$ & $2.46\times10^{12}$ \\
            $2\times10^{2}$ & $4.55\times10^{2}$ & $1.57\times10^{14}$ \\
            $5\times10^{2}$ & $1.14\times10^{3}$ & $3.84\times10^{16}$ \\
            $1\times10^{3}$ & $2.28\times10^{3}$ & $2.46\times10^{18}$ \\
            $2\times10^{3}$ & $4.54\times10^{3}$ & $1.57\times10^{20}$ \\
            $5\times10^{3}$ & $1.14\times10^{4}$ & $3.84\times10^{22}$ \\
            $1\times10^{4}$ & $2.28\times10^{4}$ & $2.46\times10^{24}$ \\ \hline
\end{tabular}
\end{table}

\begin{table}[ht]
\caption{List of $\nu _{p}$ and $(\nu $f$_\nu )_{p}$ for the synchrotron
radiation}
\begin{tabular}{lll}
\hline\hline
$\Gamma $ & $\nu _{p}/\nu _{0,S}$ & $(\nu $f$_{\nu ,S})_{p}$ \\ \hline
            $1\times10^{0}$ & $3.68\times10^{3}$ & $1.68\times10^{4}$ \\
            $2\times10^{0}$ & $1.13\times10^{4}$ & $1.07\times10^{7}$ \\
            $5\times10^{0}$ & $2.99\times10^{4}$ & $3.19\times10^{9}$ \\
            $1\times10^{1}$ & $6.00\times10^{4}$ & $2.10\times10^{11}$ \\
            $2\times10^{1}$ & $1.21\times10^{5}$ & $1.35\times10^{13}$ \\
            $5\times10^{1}$ & $3.00\times10^{5}$ & $3.31\times10^{15}$ \\
            $1\times10^{2}$ & $6.03\times10^{5}$ & $2.12\times10^{17}$ \\
            $2\times10^{2}$ & $1.20\times10^{6}$ & $1.36\times10^{19}$ \\
            $5\times10^{2}$ & $3.01\times10^{6}$ & $3.31\times10^{21}$ \\
            $1\times10^{3}$ & $6.01\times10^{6}$ & $2.12\times10^{23}$ \\
            $2\times10^{3}$ & $1.20\times10^{7}$ & $1.36\times10^{25}$ \\
            $5\times10^{3}$ & $3.01\times10^{7}$ & $3.31\times10^{27}$ \\
            $1\times10^{4}$ & $6.00\times10^{7}$ & $2.12\times10^{29}$ \\ \hline
\end{tabular}
\end{table}

\begin{table}[ht]
\caption{List of $\nu _{p}$ and $(\nu $f$_\nu )_{p}$ for the radiation of
the GRB model}
\begin{tabular}{lll}
\hline\hline
$\Gamma $ & $\nu _{p}/\nu _{0,p}$ & $(\nu $f$_{\nu ,G})_{p}$ \\ \hline
            $1\times10^{0}$ & $9.99\times10^{-1}$ & $1.84\times10^{-1}$ \\
            $2\times10^{0}$ & $3.13\times10^{0}$ & $1.17\times10^{2}$ \\
            $5\times10^{0}$ & $8.28\times10^{0}$ & $3.50\times10^{4}$ \\
            $1\times10^{1}$ & $1.66\times10^{1}$ & $2.30\times10^{6}$ \\
            $2\times10^{1}$ & $3.34\times10^{1}$ & $1.48\times10^{8}$ \\
            $5\times10^{1}$ & $8.33\times10^{1}$ & $3.62\times10^{10}$ \\
            $1\times10^{2}$ & $1.67\times10^{2}$ & $2.32\times10^{12}$ \\
            $2\times10^{2}$ & $3.34\times10^{2}$ & $1.48\times10^{14}$ \\
            $5\times10^{2}$ & $8.33\times10^{2}$ & $3.62\times10^{16}$ \\
            $1\times10^{3}$ & $1.67\times10^{3}$ & $2.32\times10^{18}$ \\
            $2\times10^{3}$ & $3.34\times10^{3}$ & $1.48\times10^{20}$ \\
            $5\times10^{3}$ & $8.33\times10^{3}$ & $3.62\times10^{22}$ \\
            $1\times10^{4}$ & $1.67\times10^{4}$ & $2.32\times10^{24}$ \\ \hline
\end{tabular}
\end{table}

We find from these tables that $\nu _p$ as well as $(\nu f_\nu
)_p$ always rise with increasing of $\Gamma $ (see Tables 1--4).
The distribution of $\nu _p$ (or $E_p$) was once proposed
(\cite{Br98}) to scale as the bulk Lorentz factor $\Gamma $. This
proposal would be valid for most cases, especially when the
expansion is not extremely large.

 \begin{figure}

 \vbox to 3.0in{\rule{0pt}{3.0in}} \includegraphics{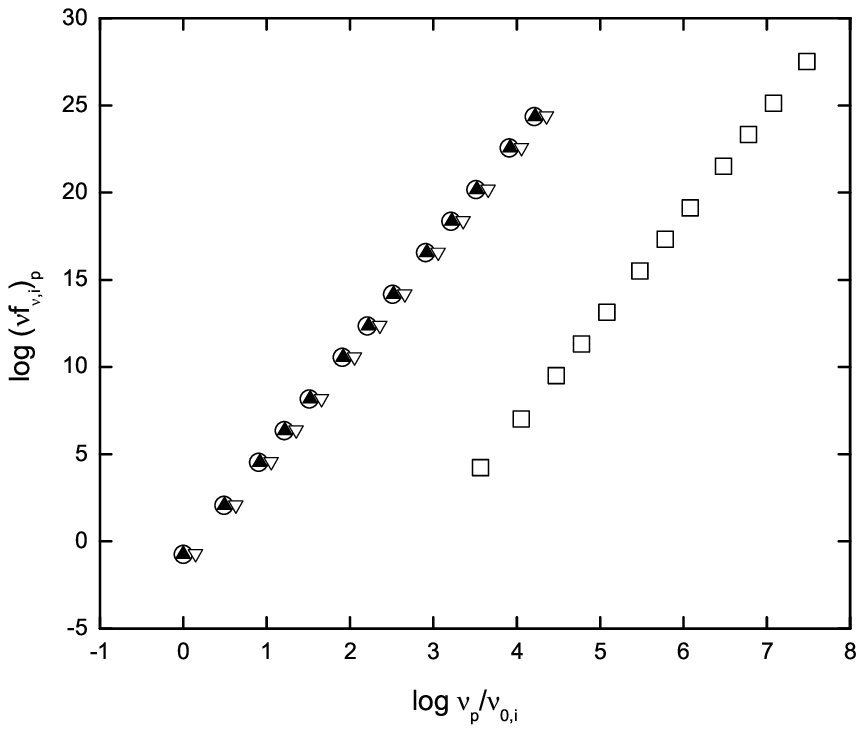}

      \caption{The plots of log$(\nu f_{\nu})_{p}- log\nu_{p}$ for
the bremsstrahlung (open circle), Comptonized (open triangle),
synchrotron (open square), and the GRB model (filled triangle)
radiation, where log$(\nu f_{\nu,i})_{p}$ represents log$(\nu
f_{\nu,B})_{p}$, log$(\nu f_{\nu,C})_{p}$, log$(\nu
f_{\nu,S})_{p}$ and log$(\nu f_{\nu,G})_{p}$, while
log$(\nu_{p}/\nu_{0,i})$ represents log$(\nu_{p}/\nu_{0,B})$,
log$(\nu_{p}/\nu_{0,C})$, log$(\nu_{p}/\nu_{0,S})$ and
log$(\nu_{p}/\nu_{0,P})$. The data are taken from Tables 1-4.
       }
       \label{Fig5}
   \end{figure}

To obtain an intuitive view of the relation between the two
elements, we present Fig. 5. It displays the plots of $\log (\nu
f_\nu )_p-\log \nu _p$ for the bremsstrahlung, Comptonized,
synchrotron, and the GRB model radiations considered above,
respectively, where $\nu _p$ is scaled respectively to the typical
frequencies adopted in Figs. 1--4 and Tables 1--4 for the
corresponding models. The figure shows clearly that the two
elements are obviously correlated. Indeed, it was discovered that
the mean peak energies of gamma-ray burst spectra are correlated
with intensity (\cite{Ma95}).

\section{Change of the spectral shape}

In this section, we investigate the change of the spectral shape caused by
the expansion of fireballs. A direct comparison between the rest frame
radiation and the expected observational radiation is made to show the
change. This is realized by adjusting the typical frequencies and the
magnitudes of the rest frame radiations, which are used for comparison, of
the corresponding models so that the values of $(\nu f_\nu )_p$ as well as $%
\nu _p$ for both the compared rest frame spectrum and the expected
observational spectrum are the same.

\begin{figure}[tbp]
\vbox to 3.0in{\rule{0pt}{3.0in}} \includegraphics{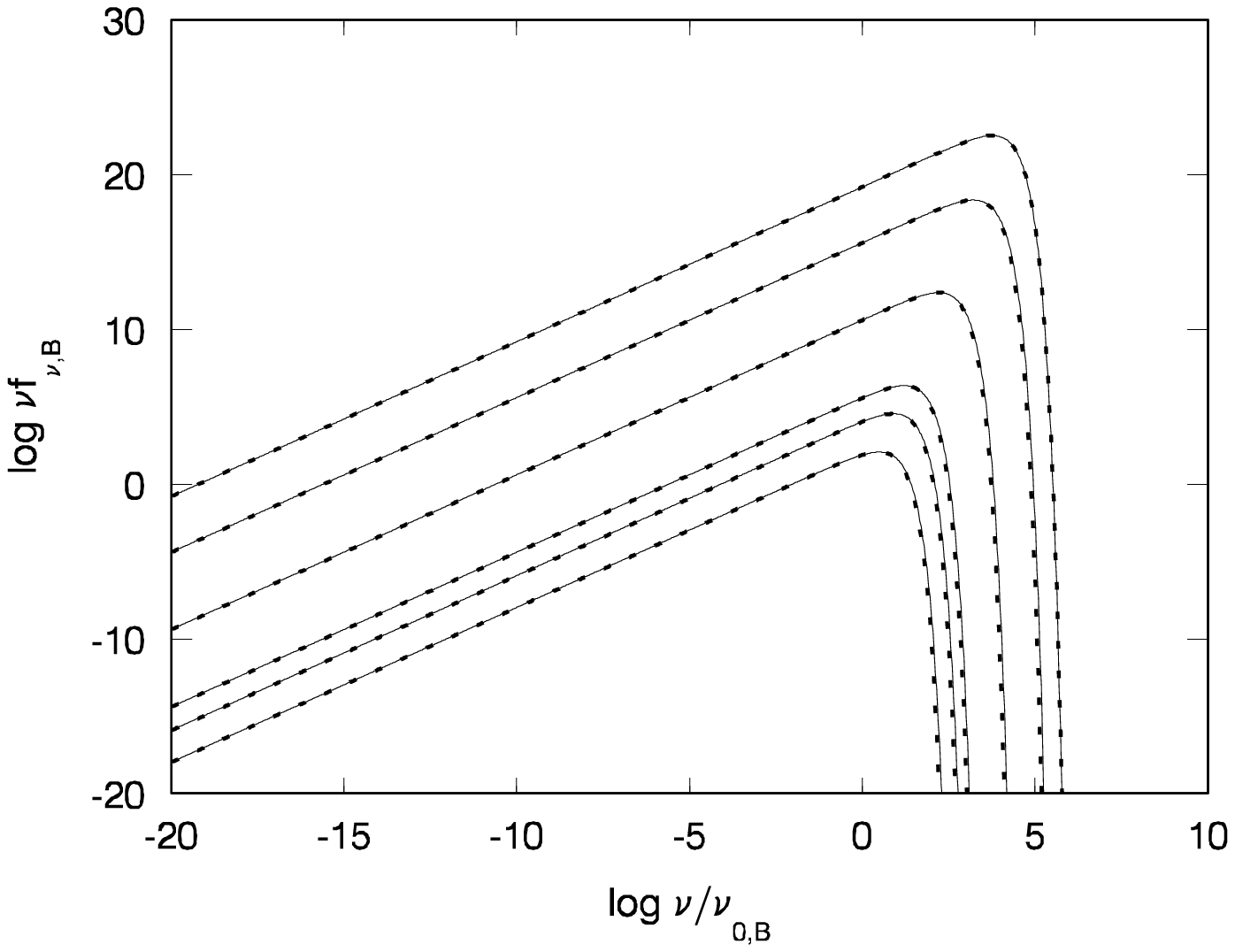} \caption{The
plot for comparing the shapes of the expected spectrum of a
fireball and the corresponding rest frame spectrum in the case of
the bremsstrahlung radiation. The solid lines from the bottom to
the top stand for the expected spectra corresponding to $\Gamma =$
2, 5, 10, 100, 1000, and 10000, respectively, while the dotted
lines from the bottom to the top are the compared rest frame
spectra for the corresponding, from the bottom to the top,
expected spectra. } \label{Fig6}
\end{figure}

\begin{figure}[tbp]
\vbox to 3.0in{\rule{0pt}{3.0in}} \includegraphics{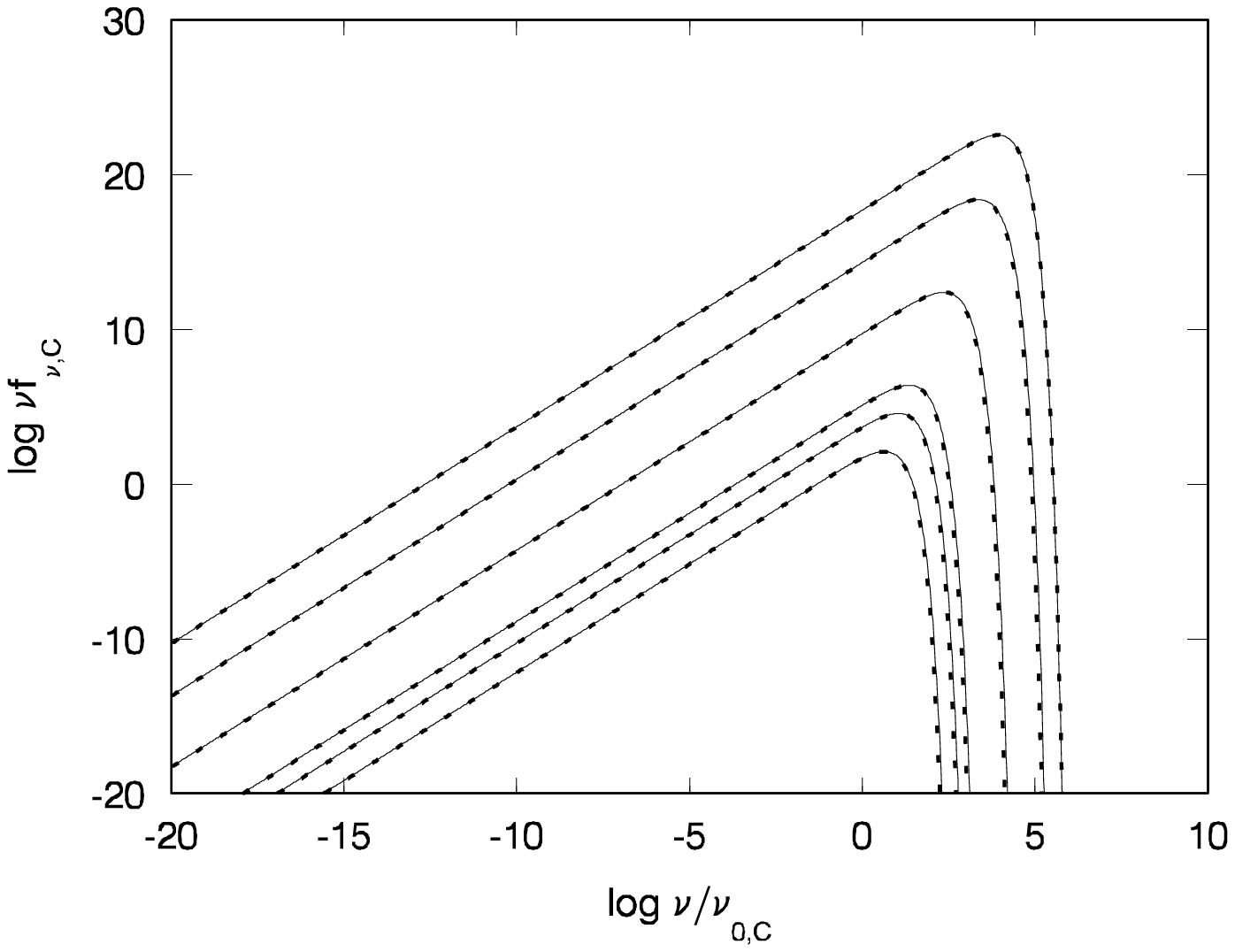} \caption{The
plot for comparing the shapes of the expected spectrum of a
fireball and the corresponding rest frame spectrum in the case of
the Comptonized radiation. The symbols are the same as those in
Fig. 6. } \label{Fig7}
\end{figure}

\begin{figure}[tbp]
\vbox to 3.0in{\rule{0pt}{3.0in}} \includegraphics{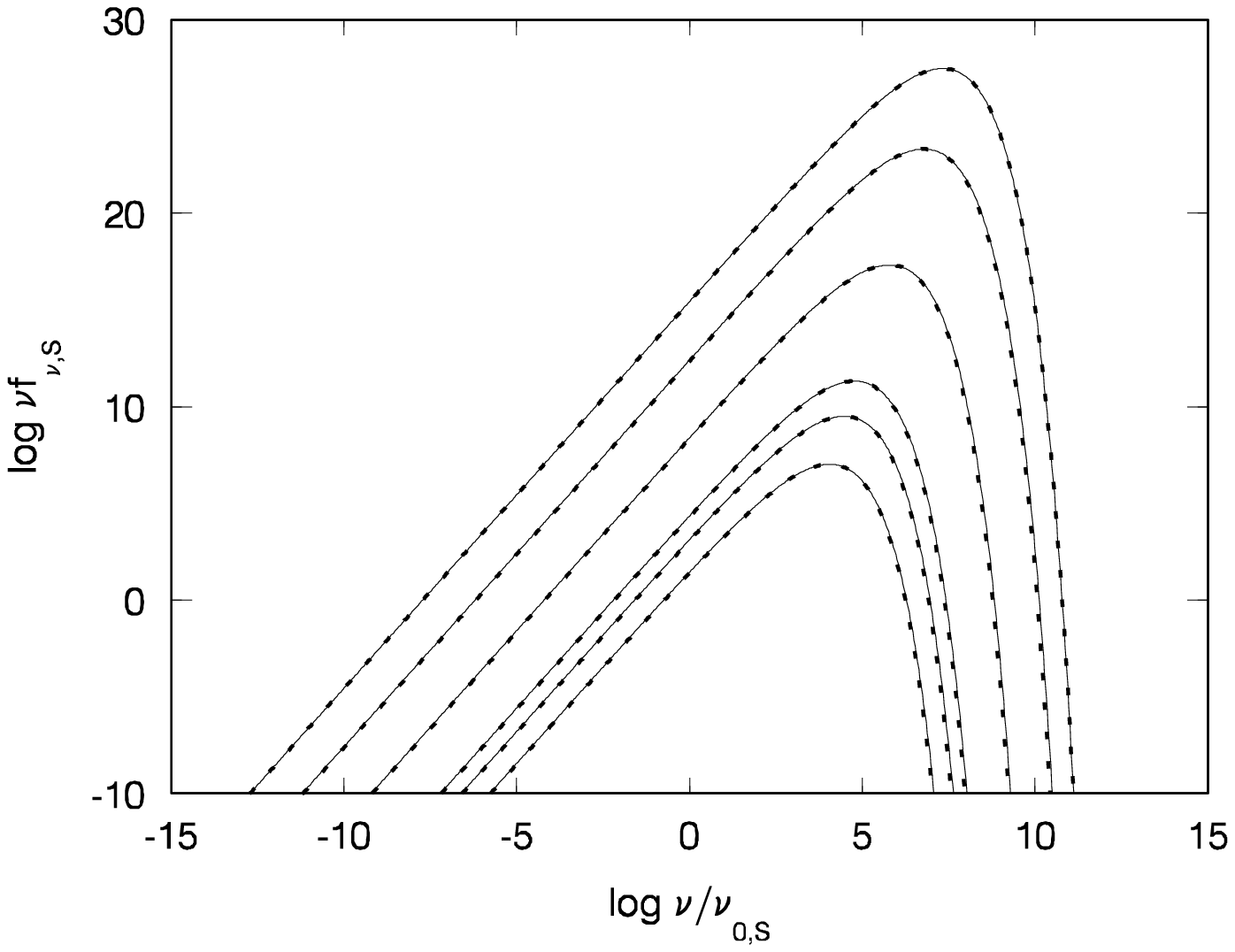} \caption{The
plot for comparing the shapes of the expected spectrum of a
fireball and the corresponding rest frame spectrum in the case of
the synchrotron radiation. The symbols are the same as those in
Fig. 6. } \label{Fig8}
\end{figure}

\begin{figure}[tbp]
\vbox to 3.0in{\rule{0pt}{3.0in}} \includegraphics{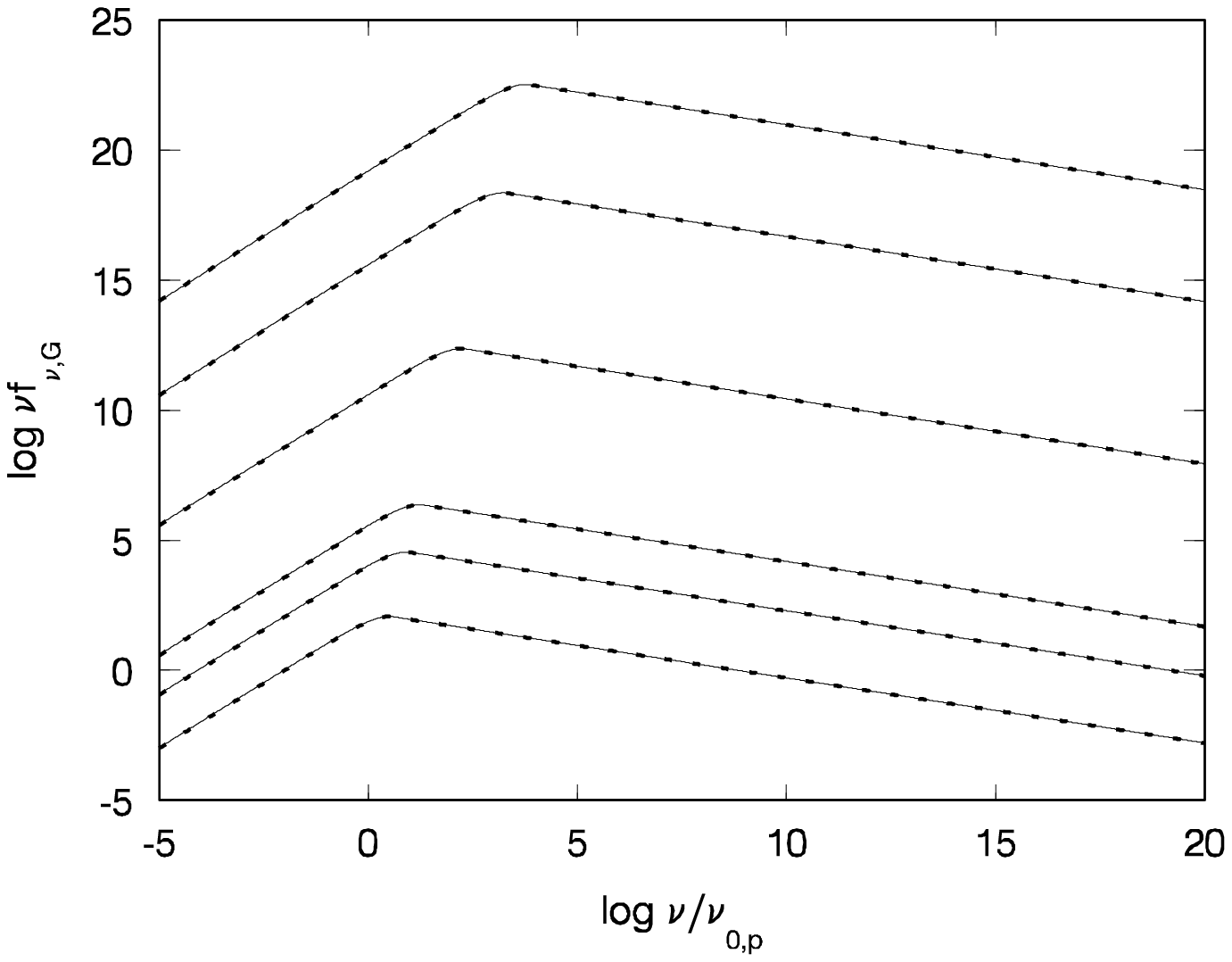} \caption{The
plot for comparing the shapes of the expected spectrum of a
fireball and the corresponding rest frame spectrum in the case of
the radiation bearing the GRB form. The symbols are the same as
those in Fig. 6. } \label{Fig9}
\end{figure}

Shown in Figs. 6--9 are the same plots of Figs. 1--4, where the expected
observational spectra as well as the corresponding compared rest frame
spectra are presented.

These figures show that when the expansion speed is not extremely
large (say, $\Gamma \leq 1000$), the expected spectra become
slightly widened and are less steep at higher frequency bands for
the bremsstrahlung, Comptonized, and synchrotron radiations, and
at lower frequency bands for the GRB model radiation. The shapes
of the expected spectra are almost the same as the rest frame
ones. Therefore, for constant radiations, when a certain form can
well describe the observed spectrum at a given observation time,
it would also be able to represent the observed spectrum at other
observation time. In particular, when we observe a decrease of the
temperature (say, $T_B$ in the bremsstrahlung or $T_C$ in the
Comptonized radiation, see \cite{Sc94}), we would expect a
decrease of the intensity as well due the decrease of the
expansion speed (which causes the observed decrease of the
temperature), as long as the radiation is constant.

\section{Enhancement of the flux}

We note that the expected flux of an expanding fireball would vary
with both frequency and the expansion speed. Different from those
figures shown in section 3, we present here plots of $\log (\nu
f_\nu )-\log \Gamma $ at some particular frequencies for the
radiations considered above, which would plainly show the
enhancement of fluxes caused by the expansion of fireballs. They
are Figs. 10--13. Figs. 10--12 show that, for constant radiations
of the bremsstrahlung, Comptonized, and synchrotron mechanisms,
when the expansion speed decreases steadily, fluxes at higher
frequencies would decrease very rapidly and later might become
undetectable, while those at lower frequencies would also decrease
but in a much slow {\bf manner}. But this phenomenon would be rare
because the number of fireballs with the expansion speed that
large would be small. From Fig. 13 one finds that, for a constant
radiation of the GRB form, when the expansion speed decreases
steadily, fluxes at any (lower or higher) frequencies would
decrease in almost the same {\bf manner}.

\begin{figure}[tbp]
\vbox to 3.0in{\rule{0pt}{3.0in}} \includegraphics{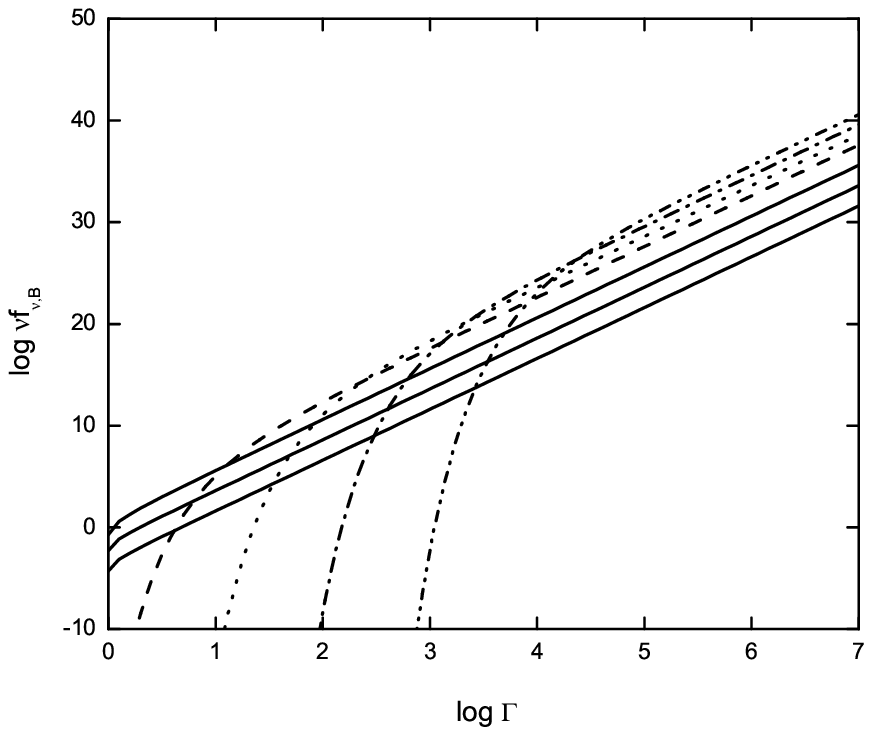}
 \caption{The
plot of log($\nu $f$_{\nu ,B}$) --- log$\Gamma $ for the
bremsstrahlung radiation at various frequencies [the solid lines
from the bottom to the top: log($\nu /\nu _{0,B}$) = --4, --2, 0;
the dashed line: log($\nu /\nu _{0,B}$) = 2; the dotted line:
log($\nu /\nu _{0,B}$) = 3; the dashed and dotted line: log($\nu
/\nu _{0,B}$) = 4; the short and long dashed line: log($\nu /\nu
_{0,B}$) = 5]. } \label{Fig10}
\end{figure}

\begin{figure}[tbp]
\vbox to 3.0in{\rule{0pt}{3.0in}} \includegraphics{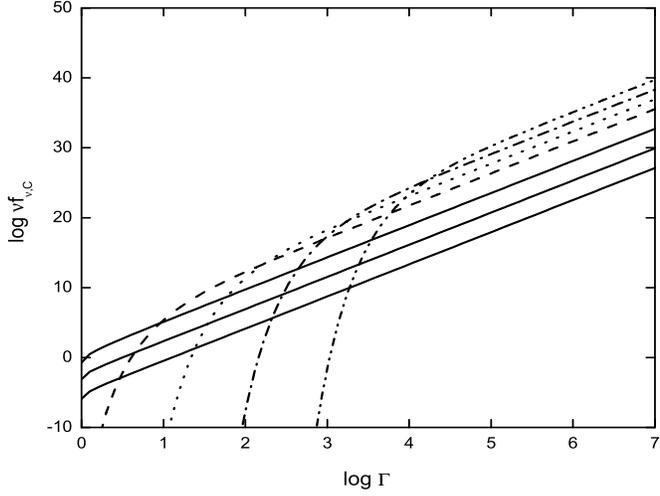} \caption{The
plot of log($\nu $f$_{\nu ,C}$) --- log$\Gamma $ for the
Comptonized radiation at various frequencies [the solid lines from
the bottom to the top: log($\nu /\nu _{0,C}$) = --4, --2, 0; the
dashed line: log($\nu /\nu _{0,C}$) = 2; the dotted line: log($\nu
/\nu _{0,C}$) = 3; the dashed and dotted line: log($\nu /\nu
_{0,C}$) = 4; the short and long dashed line: log($\nu /\nu
_{0,C}$) = 5]. } \label{Fig11}
\end{figure}

\begin{figure}[tbp]
\vbox to 3.0in{\rule{0pt}{3.0in}} \includegraphics{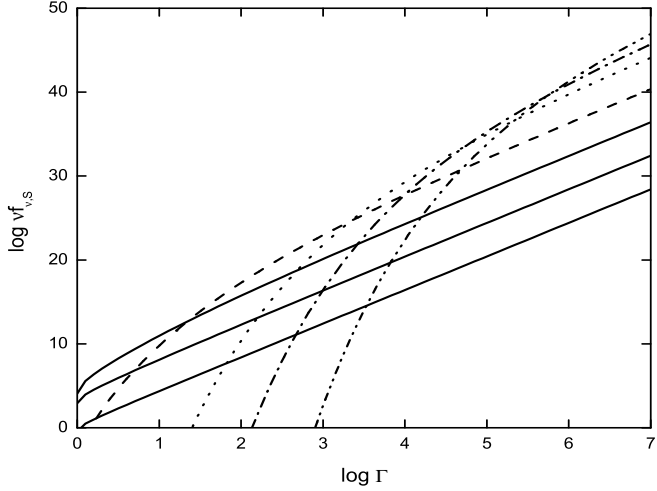} \caption{The
plot of log($\nu $f$_{\nu ,S}$) --- log$\Gamma $ for the
synchrotron radiation at various frequencies [the solid lines from
the
bottom to the top: log($\nu /\nu _{0,S}$) = 0, 2, 4; the dashed line: log($%
\nu /\nu _{0,S}$) = 6; the dotted line: log($\nu /\nu _{0,S}$) = 8; the
dashed and dotted line: log($\nu /\nu _{0,S}$) = 9; the short and long
dashed line: log($\nu /\nu _{0,S}$) = 10]. }
\label{Fig12}
\end{figure}

\begin{figure}[tbp]
\vbox to 3.0in{\rule{0pt}{3.0in}} \includegraphics{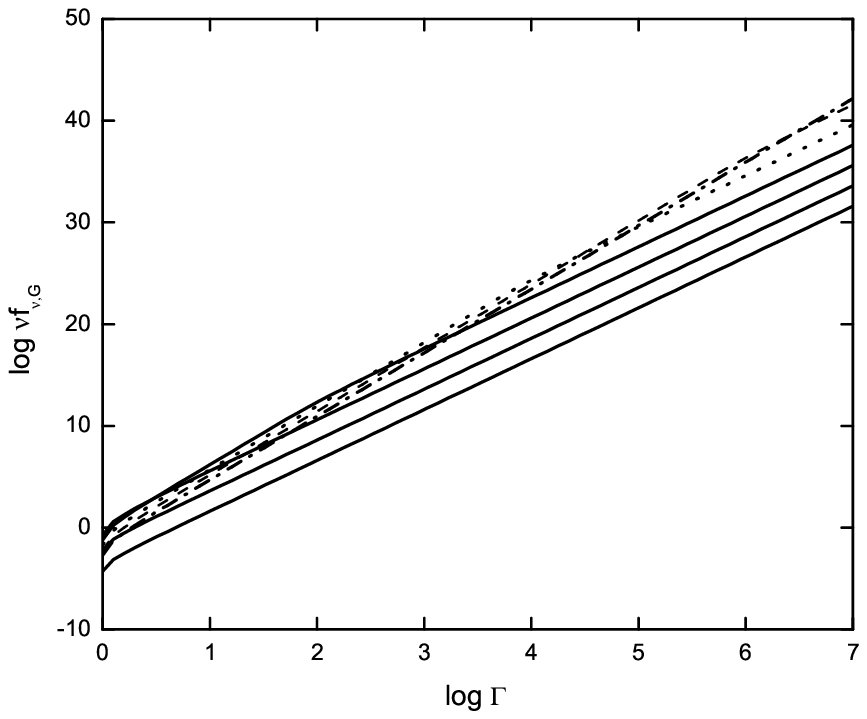} \caption{The
plot of log($\nu $f$_{\nu ,G}$) --- log$\Gamma $ for the radiation
bearing the GRB form at various frequencies [the solid lines from
the bottom to the top: log($\nu /\nu _{0,p}$) = --4, --2, 0, 2;
the dotted line: log($\nu /\nu _{0,p}$) = 4; the dashed line:
log($\nu /\nu _{0,p}$) = 6; the dashed and dotted line: log($\nu
/\nu _{0,p}$) = 8]. } \label{Fig13}
\end{figure}

These figures also show that, the expansion speed plays a very important
role in the enhancement of fluxes: when the value of $\Gamma $ becomes 10
times larger, the flux would be several orders of magnitude larger. The
expansion speed of fireballs would determine relative magnitudes of their
fluxes observed and thus might determine if a gamma-ray source is
detectable. Note that $(1+z)^2$ is the main factor of the cosmological
effect. That suggests that the effect caused by the expansion speed of
fireballs is really very large and would be the fundamental factor of the
enhancement of the flux of the objects.

\section{Conclusions}

In this paper, we investigate how the Doppler effect in the fireball
framework plays a role on the spectrum of gamma-ray bursts.

We find that the shape of the expected spectrum of an expanding
fireball remains almost the same as that of the corresponding rest
frame spectrum for constant radiations of the bremsstrahlung,
Comptonized, and synchrotron mechanisms as well as for that of the
GRB model. However, as the fireball expands, the peak of the
spectrum would shift to a higher frequency band and the flux over
the entire frequency range would be amplified. The study reveals
that the {\bf peak flux spectrum} $(\nu f_\nu )_p$ and the peak
frequency $\nu _p$ are obviously correlated, meeting what was
discovered recently (\cite{Ma95}). {\bf The expansion speed plays
a very
important role in the enhancement of fluxes: when the value of }$\Gamma $%
{\bf \ becomes 10 times larger, the flux would be several orders
of magnitude larger. Even the expansion speed is not very large
(say }$\Gamma =2 ${\bf ), a dim and undetectable X-ray rest frame
radiation might become an observable gamma-ray source. The
expansion speed of fireballs is a fundamental factor of the
enhancement of the flux of GRBs.}

The study shows that, for constant radiations of the
bremsstrahlung, Comptonized, and synchrotron mechanisms, when the
expansion speed decreases steadily, fluxes at higher frequencies
would decrease very rapidly and later might become undetectable,
while those at lower frequencies would also decrease but in a much
slow {\bf manner}. However, for a constant radiation of the GRB
form, when the expansion speed decreases steadily, fluxes at any
(lower or higher) frequencies would decrease in almost the same
{\bf manner}.
\vspace{8mm}

It is my great pleasure to thank Prof. M. J. Rees for his helpful
suggestions and discussion. Parts of this work were done when I
visited Institute of Astronomy, University of Cambridge and
Astrophysics Research Institute, Liverpool John Moores University.
Thanks are also given to these institutes. This work was supported
by the Special Funds for Major State Basic Research Projects,
National Natural Science Foundation of China.

\appendix

\section{Detailed derivation of the formula}

We concern a fireball expanding at a constant velocity $v=\beta c$ {\bf %
(where }$\mathbf{c}${\bf \ is the speed of light)} and adopt a
spherical coordinate system with its origin being sited at the
center of the fireball and its axis being the line of sight.
Consider radiation from the rest frame differential surface,
$ds_{0,\theta ,\varphi }$, of the fireball at proper time
$t_{0,\theta }$, {\bf where }$\theta ${\bf \ denotes the angle to
the line of sight and }$\varphi ${\bf \ denotes the other angular
coordinate of the fireball surface}. Let $ds_{\theta ,\varphi }$
be the corresponding differential surface resting on the observer
framework,
coinciding with $ds_{0,\theta ,\varphi }$ at $t_{0,\theta }$. Obviously, $%
ds_{0,\theta ,\varphi }$ moves at velocity $v$ relative to
$ds_{\theta ,\varphi }$.

Let $t_\theta $ be the corresponding coordinate time when $ds_{\theta
,\varphi }$ coincides with $ds_{0,\theta ,\varphi }$ at $t_{0,\theta }$.
According to the theory of special relativity, $t_\theta $ and $t_{0,\theta }
$ are related by
\begin{equation}
t_\theta -t_c=\Gamma (t_{0,\theta }-t_{0,c}),
\end{equation}
where
$t_c$ and $t_{0,c}$ are constants (here we assign $t_\theta =t_c$
when $t_{0,\theta }=t_{0,c}$), and $\Gamma $ is the Lorentz factor
of the fireball.

The area of $ds_{\theta ,\varphi }$ is
\begin{equation}
ds_{\theta ,\varphi }=R^2(t_\theta )\sin \theta d\theta d\varphi ,
\end{equation}
where $R(t_\theta )$ is the radius of the fireball at $t_\theta $. The
radius follows
\begin{equation}
R(t_\theta )=\beta c(t_\theta -t_c)+R_c,
\end{equation}
where $R_c$ is the radius at time $t_\theta =t_c$. As assigned above, $%
t_\theta $ and $t_{0,\theta }$ correspond to the same moment, thus the
radius can also be expressed as
\begin{equation}
\ R_0(t_{0,\theta })=R[t_\theta (t_{0,\theta })]=\Gamma \beta c(t_{0,\theta
}-t_{0,c})+R_c,
\end{equation}
where equation (A1) is applied.

Let us consider an observation within the small intervals $t$ --- $t+dt$ and
$\nu $ --- $\nu +d\nu $ carried out by an observer with a detector $%
\triangle s_{ob}$ at a distance $D$ ($D$ is the distance between the
observer and the center of the fireball), where $D\gg R(t_\theta )$. Suppose
radiation from $ds_{0,\theta ,\varphi }$ arriving the observer within the
above observation intervals is emitted within the proper time interval $%
t_{0,\theta }$ --- $t_{0,\theta }+dt_{0,\theta }$ and the rest frame
frequency interval $\nu _{0,\theta }$ --- $\nu _{0,\theta }+d\nu _{0,\theta }
$. According to the Doppler effect, $\nu $ and $\nu _{0,\theta }$ are
related by
\begin{equation}
\nu =\frac{\nu _{0,\theta }}{\Gamma (1-\beta \cos \theta )}.
\end{equation}

Considering the travelling of light from the fireball to the observer, one
would come to
\begin{equation}
c(t-t_\theta )=D-R(t_\theta )\cos \theta ,
\end{equation}
(note that the cosmological effect is ignored). Combining (A1), (A3) and
(A6) yields
\begin{equation}
t_\theta =\frac{t-D/c+(R_c/c-\beta t_c)\cos \theta }{1-\beta \cos \theta }
\end{equation}
and
\begin{equation}
t_{0,\theta }=\frac{t-t_c-D/c+(R_c/c)\cos \theta }{\Gamma (1-\beta \cos
\theta )}+t_{0,c}.
\end{equation}
The radius of the fireball then can be written as
\begin{equation}
R[t_\theta (t)]=R_0[t_{0,\theta }(t)]=\frac{\widetilde{R}(t)}{1-\beta \cos
\theta },
\end{equation}
with
\begin{equation}
\widetilde{R}(t)=\beta [c(t-t_c)-D]+R_c.
\end{equation}

Suppose photons, which are emitted from $ds_{0,\theta ,\varphi }$ within
proper time interval $t_{0,\theta }$ --- $t_{0,\theta }+dt_{0,\theta }$ and
then reach the observer within $t$ --- $t+dt$, pass through $ds_{\theta
,\varphi }$ within coordinate time interval $t_\theta $ --- $t_\theta
+dt_{\theta ,s}$. Since both $ds_{\theta ,\varphi }$ and the observer rest
on the same framework, it would be held that $dt_{\theta ,s}=dt$ (when the
cosmological effect is ignored). Of course, the frequency interval for the
photons measured by both $ds_{\theta ,\varphi }$ and the observer must be
the same: $\nu $ --- $\nu +d\nu $. Suppose the radiation concerned is
independent of directions. Then in the view of $ds_{\theta ,\varphi }$
(which is also the view of the observer), the amount of energy emitted from $%
ds_{0,\theta ,\varphi }$ within $t_{0,\theta }$ --- $t_{0,\theta
}+dt_{0,\theta }$ and $\nu _{0,\theta }$ --- $\nu _{0,\theta }+d\nu
_{0,\theta }$ (which would pass through $ds_{\theta ,\varphi }$ within $%
t_\theta $ --- $t_\theta +dt$ and be measured within $\nu $ --- $\nu +d\nu $%
) towards the observer would be
\begin{equation}
dE_{\theta ,\varphi }=I_\nu (t_\theta ,\nu )\cos \theta ds_{\theta ,\varphi
}d\omega d\nu dt,
\end{equation}
where $I_\nu (t_\theta ,\nu )$ is the intensity of radiation measured by $%
ds_{\theta ,\varphi }$ or by the observer and $d\omega $ is the solid angle
of $\triangle s_{ob}$ with respect to the fireball, which is
\begin{equation}
d\omega =\frac{\triangle s_{ob}}{D^2}.
\end{equation}
(It is clear that any elements of emission from the fireball are independent
of $\varphi $ due to the symmetric nature of the object.) Thus,
\begin{equation}
dE_{\theta ,\varphi }=\frac{\triangle s_{ob}d\nu dtR^2(t_\theta )I_\nu
(t_\theta ,\nu )\cos \theta \sin \theta d\theta d\varphi }{D^2},
\end{equation}
where (A2) and (A12) are applied.

The total amount of energy emitted from the whole fireball surface detected
by the observer within the above observation intervals is an integral of $%
dE_{\theta ,\varphi }$ over that area, which is
\begin{equation}
dE=\frac{2\pi \triangle s_{ob}d\nu dt}{D^2}\int_{\theta _{\min }}^{\theta
_{\max }}R^2(t_\theta )I_\nu (t_\theta ,\nu )\cos \theta \sin \theta d\theta
,
\end{equation}
where $\theta _{\min }$ and $\theta _{\max }$ are determined by the fireball
surface itself together with the emitted ranges of $t_{0,\theta }$ and $\nu
_{0,\theta }$. Thus, the expected flux would be
\begin{equation}
f_\nu (t)=\frac{2\pi }{D^2}\int_{\theta _{\min }}^{\theta _{\max
}}R^2(t_\theta )I_\nu (t_\theta ,\nu )\cos \theta \sin \theta d\theta .
\end{equation}
It is well known that the observer frame intensity $I_\nu (t_\theta ,\nu )$
is related to the rest frame intensity $I_{0,\nu }(t_{0,\theta },\nu
_{0,\theta })$ by
\begin{equation}
I_\nu (t_\theta ,\nu )=(\frac \nu {\nu _{0,\theta }})^3I_{0,\nu
}(t_{0,\theta },\nu _{0,\theta }).
\end{equation}
The flux then can be written as
\begin{equation}
f_\nu (t)=\frac{2\pi }{D^2\Gamma ^3}\int_{\theta _{\min }}^{\theta _{\max }}%
\frac{R_0^2(t_{0,\theta })I_{0,\nu }(t_{0,\theta },\nu _{0,\theta })\cos
\theta \sin \theta }{(1-\beta cos\theta )^3}d\theta ,
\end{equation}
where (A4) and (A5) are applied.

Now let us find out how $\theta _{\min }$ and $\theta _{\max }$ are
determined. We are aware that the range of $\theta $ of the visible fireball
surface is
\begin{equation}
0\leq \theta \leq \pi /2.
\end{equation}
Within this range, suppose the emitted ranges of $t_{0,\theta }$ and $\nu
_{0,\theta }$ constrain $\theta $ by
\begin{equation}
\theta _{t,\min }\leq \theta \leq \theta _{t,\max }
\end{equation}
and
\begin{equation}
\theta _{\nu ,\min }\leq \theta \leq \theta _{\nu ,\max },
\end{equation}
respectively. Then when the following condition
\begin{equation}
\max \{\theta _{t,\min },\theta _{\nu ,\min }\}<\min \{\theta _{t,\max
},\theta _{\nu ,\max }\}
\end{equation}
is satisfied, $\theta _{\min }$ and $\theta _{\max }$ would be obtained by
\begin{equation}
\theta _{\min }=\max \{\theta _{t,\min },\theta _{\nu ,\min }\}
\end{equation}
and
\begin{equation}
\theta _{\max }=\min \{\theta _{t,\max },\theta _{\nu ,\max }\},
\end{equation}
respectively.

Let the emitted ranges of $t_{0,\theta }$ and $\nu _{0,\theta }$ be
\begin{equation}
t_{0,\min }\leq t_{0,\theta }\leq t_{0,\max }
\end{equation}
and
\begin{equation}
\nu _{0,\min }\leq \nu _{0,\theta }\leq \nu _{0,\max },
\end{equation}
respectively. From (A8) and (A24) one gets
\begin{equation}
\begin{tabular}{c}
$\theta _{t,\min }=\cos ^{-1}\left( \min \{1,\frac{\Gamma c(t_{0,\max
}-t_{0,c})+D-c(t-t_c)}{\Gamma \beta c(t_{0,\max }-t_{0,c})+R_c}\}\right) $
\\
$\left( 0\leq \frac{\Gamma c(t_{0,\max }-t_{0,c})+D-c(t-t_c)}{\Gamma \beta
c(t_{0,\max }-t_{0,c})+R_c}\right) $%
\end{tabular}
\end{equation}
and
\begin{equation}
\begin{tabular}{c}
$\theta _{t,\max }=\cos ^{-1}\left( \max \{0,\frac{\Gamma c(t_{0,\min
}-t_{0,c})+D-c(t-t_c)}{\Gamma \beta c(t_{0,\min }-t_{0,c})+R_c}\}\right) $
\\
$\left( \frac{\Gamma c(t_{0,\min }-t_{0,c})+D-c(t-t_c)}{\Gamma \beta
c(t_{0,\min }-t_{0,c})+R_c}\leq 1\right) .$%
\end{tabular}
\end{equation}
For $\beta >0$, one can obtain the following from (A5) and (A25):
\begin{equation}
\begin{tabular}{c}
$\theta _{\nu ,\min }=\cos ^{-1}\left( \min \{1,\frac 1\beta (1-\frac{\nu
_{0,\min }}{\Gamma \nu })\}\right) $ \\
$\left( 0\leq \frac 1\beta (1-\frac{\nu _{0,\min }}{\Gamma \nu });0<\beta
\right) ,$%
\end{tabular}
\end{equation}
\begin{equation}
\begin{tabular}{c}
$\theta _{\nu ,\max }=\cos ^{-1}\left( \max \{0,\frac 1\beta (1-\frac{\nu
_{0,\max }}{\Gamma \nu })\}\right) $ \\
$\left( \frac 1\beta (1-\frac{\nu _{0,\max }}{\Gamma \nu })\leq 1;0<\beta
\right) .$%
\end{tabular}
\end{equation}

\section{Condition for $\theta _{t,\min }=0$ and $\theta _{t,\max }=\pi /2$}

Here we show the conditions for retaining $\theta _{t,\min }=0$ and $\theta
_{t,\max }=\pi /2$.

Since $R(t_\theta )>0$, from (A4) and (A8) we find that
\begin{equation}
\frac{\beta c(t-t_c)-\beta D+R_c}{\Gamma \beta c(1-\beta \cos \theta )}>0.
\end{equation}
Then from (A8) we obtain
\begin{equation}
\frac{dt_{0,\theta }}{d\cos \theta }=\frac{\beta c(t-t_c)-\beta D+R_c}{%
\Gamma c(1-\beta \cos \theta )^2}>0.
\end{equation}
It suggests that
\begin{equation}
t_{0,\pi /2}\leq t_{0,\theta }\leq t_{0,0}.
\end{equation}
From (A24) and (B3) one finds that, when
\begin{equation}
t_{0,\min }\leq t_{0,\pi /2}
\end{equation}
and
\begin{equation}
t_{0,\max }\geq t_{0,0},
\end{equation}
one would get $\theta _{t,\min }=0$ and $\theta _{t,\max }=\pi /2$. Applying
(A8) leads to
\begin{equation}
t_{0,\min }\leq \frac{c(t-t_c)-D}{\Gamma c}+t_{0,c}
\end{equation}
and
\begin{equation}
t_{0,\max }\geq \frac{c(t-t_c)-D+R_c}{\Gamma c(1-\beta )}+t_{0,c}.
\end{equation}

\clearpage

\end{document}